\documentstyle[12pt]{article}
\input{psfig.sty}
\begin{document}
%\received{1997 October 17}
\date{}
\title{Joule heating and the thermal evolution
       of old neutron stars}
\author{Juan A. Miralles$^{1)}$, Vadim Urpin$^{2,3)}$ and 
Denis Konenkov$^{3)}$\\
      $^{1)}$ Departament d'Astronomia i Astrofisica, Universitat de 
      Val\`encia, \\ 
      Dr. Moliner 50, E-46100 Burjassot (Valencia), Spain \\
      $^{2)}$ Department of Mathematics, University of Newcastle \\
            Newcastle upon Tyne NE1 7RU, UK \\
      $^{3)}$ A.F.Ioffe Institute of Physics and Technology, \\
        194021 St.Petersburg, Russia }

\maketitle

\begin{abstract}

We consider Joule heating caused by dissipation of the magnetic field
in the neutron star crust. This mechanism may be efficient in maintaining    
a relatively high surface temperature in very old neutron stars.
Calculations of the thermal evolution show that, at the late evolutionary
stage ($t \geq 10$ Myr), the luminosity of the neutron star is 
approximately equal to the energy released due to the field dissipation
and is practically independent of the atmosphere models. At this stage, 
the surface temperature can be of the order of $3 \times 10^{4} - 10^{5}$K. 
Joule heating can maintain this high temperature during extremely long 
time ($\geq 100$ Myr), comparable with the decay time of the magnetic field.
\end{abstract}

{\it Subject headings:} {magnetic fields---stars: neutron---pulsars: general}
\vspace{1cm}

\begin{center}
{\it Accepted for publication in The Astrophysical Journal}
\end{center}

\section{Introduction}

Neutron stars are very hot at birth, with temperatures well above
$10^{10}$ K. This heat is radiated away mainly by neutrinos from the
inner layers during the first million years or so (the neutrino
cooling era) and, later on, the emission of photons from the
surface dominates the cooling of the star. This photon luminosity
and its change with time depend on the properties of matter inside
the neutron star and its magnetic field. Observations of this 
radiation can thus provide important information about the state
of matter above and below nuclear density as well as about the 
magnetic field. 

The magnetic field can influence the thermal evolution of neutron
stars in different ways. This influence is probably less appreciable
during the neutrino cooling era despite the neutrino emissivities 
for some mechanisms can essentially alter in strong 
magnetic fields $\sim 10^{13}$. The effect of the magnetic field 
may be of a particular importance for neutrino processes in the inner
crust where, at some conditions, the synchrotron emission may dominate
the rate of neutrino cooling (Vidaurre et al. 1995).

The influence of the magnetic field on the cooling history is much more 
important during the photon cooling era. In the presence of a strong 
field ($\sim 10^{12} - 10^{13}$ G) the transport properties of plasma
are different compared to those in non-magnetic neutron stars.
Both the electron and radiative thermal conductivities can be affected
by the field. Generally speaking, the cooling efficiency longitudinal 
to the field lines exceeds that in the transverse direction. This change 
in the thermal conductivity can have an appreciable effect on the thermal 
evolution (see Van Riper 1991, Nomoto and Tsuruta 1987). Besides, an 
anisotropic heat transport will result in a characteristic temperature 
difference between the hot magnetic poles of the neutron star and the 
relatively cold magnetic equator (see, e.g., Schaaf 1990). 

Additional heating associated with the ohmic dissipation of currents 
may be one more important effect caused by the magnetic field. The rate 
of Joule heating depends on both the geometry of the magnetic field
and conductive properties of plasma and may be rather high for some
magnetic configurations. The total magnetic energy of the neutron star 
can probably reach $10^{43} - 10^{44}$ erg. Observational data on the 
spin and magnetic evolution of isolated pulsars provide some evidences 
that the field decay is rather slow in isolated pulsars. Thus, according 
to Narayan and Ostriker (1990) the decay time-scale is about 20 Myr. 
Bhattacharya et al. (1992) inferred even a longer decay time ($ \geq 
30-100$ Myr) from the same observational data. Nevertheless, even for 
such a slow decay the rate of Joule heating (if the decay is caused by 
ohmic dissipation) may be as large as $10^{28} - 10^{30}$ erg/s. Clearly, 
Joule heating cannot change substantially the early thermal evolution 
when the neutron star is relatively hot and its luminosity 
exceeds this value. However, the ohmic dissipation produces  
enough heat to change completely the thermal history of old neutron
stars. Assuming that all heat released due to dissipation of the 
magnetic field is transferred to the surface (that is true during the photon
cooling era) and emitted with the 
blackbody spectrum, one can obtain an estimate of the surface
temperature, $T_{s}$, required to maintain the neutron star under
the thermal equilibrium. This temperature may be as high as $3 \times 
10^{4} - 10^{5}$ K, and the neutron star can maintain this temperature
during a long time, comparable with the decay time of the magnetic
field, $\sim 30-100$ Myr, whereas $T_{s}$ has to fall down to the value 
below $10^{5}$ K after $\sim 1-3$ Myr in accordance with the so called 
standard cooling scenario. Therefore an observational and 
theoretical study of the late thermal evolution of neutron stars
may be a powerful diagnostics of their magnetic fields and can provide
an important information about the magnetic configuration and mechanisms
of its decay. 

Detection of the thermal radiation from old and close neutron stars with 
the surface temperature $\leq 10^{5}$ K has became possible only in
recent years. Thus, Becker \& Tr\"umper (1997) detected several 
middle-aged and old neutron stars in the soft X-ray band. The Hubble 
Space Telescope also detected the optical and UV thermal emission from 
few relatively old pulsars (Pavlov, Stringfellow \& Cordova 1996, 
Mignani, Caraveo \& Bignami 1997). The corresponding surface temperatures 
turn out surprisingly high compared to predictions of the standard 
cooling model. Such high temperatures can be understood only if some 
mechanisms of additional heating operate in relatively old neutron 
stars.

One possible source of heating can be caused by the frictional 
interaction of neutron superfluid with the normal matter in the inner 
crust (Shibazaki \& Lamb 1989, Umeda et al. 1993). If the inner crust
of neutron stars contains superfluid rotating faster than the rest of
the star, the differential rotation causes the frictional heat
generation. The authors found that the rate of a frictional heat
generation does not depend on specific models for the superfluid-crust
interaction.
Note that the frictional heating is independent
of the magnetic field whereas the Joule heating is strongly 
sensitive to the field strength. This difference provides a hope 
to discriminate between these heating mechanisms from observational 
data.

In the present paper, we consider the thermal evolution of a neutron
star assuming the crustal origin of its magnetic field. The models
with a crustal magnetic field turn out to be quite suitable to account 
for a wide variety of observational data on the magnetic and spin 
evolution of both isolated and entering binary systems neutron stars 
(Urpin \& Konenkov 1997, Urpin, Geppert \& Konenkov 1997). We calculate
the rate of Joule heating caused by dissipation of currents in the
neutron star crust. Incorporating the expression for Joule heating
into the numerical codes of thermal evolution, we examine the effect
of this heating on the cooling history.  In \S 2 the physical model 
adopted for our calculations is described. The results of calculations
of the thermal evolution with Joule heating are presented in \S 3.
In \S 4 we briefly summaries the results of our study.

\section{Basic equations}

We assume that the magnetic field has been generated in the neutron
star crust by some unspecified mechanism during or shortly after
neutron star formation. The evolution of such a field is controlled 
by the conductive properties of the crust. Shortly after collapse
the main fraction of the crustal material solidifies, and the evolution
of the crustal field is governed by the induction equation
without the convective term,
$$
\frac{\partial {\bf B}}{\partial t} = - \frac{c^{2}}{4 \pi}
{\bf \nabla} \times \left( \frac{1}{\sigma} {\bf \nabla}\times{\bf B}
\right) \;, \eqno(1)
%\frac{\partial \vec{B}}{\partial t} = - \frac{c^{2}}{4 \pi}
%\nabla \times \left( \frac{1}{\sigma} \nabla \times \vec{B}
%\right) \;, \eqno(1)
$$
where $\sigma$ is the conductivity. We restrict our consideration
to a dipolar field which can be described by the vector potential
${\bf A} =(0,0, A_{\varphi})$, $A_{\varphi} = S(r,t) \sin \theta /r$,
where $r$ and $\theta$ are the spherical radius and polar angle,
respectively. Then the function $S(r,t)$ obeys the equation (see,
e.g., Sang \& Chanmugam 1987)
$$
\frac{\partial^{2} S}{\partial r^{2}} - \frac{2 S}{r^{2}} =
\frac{4 \pi \sigma}{c^{2}} \frac{\partial S}{\partial t}  \eqno(2)
$$
with the boundary condition 
$$
\frac{\partial S}{\partial r} + \frac{S}{R} =0  \eqno(3)
$$
at the stellar surface $r=R$. For a field confined to the crust,
$S(r,t)$ should vanish in the deep layers. The field components in the 
interior of the star are
$$
B_{r} = \frac{2 S}{r^{2}} \cos \theta \;, \;\;\;\;\;
B_{\theta} = - \frac{\sin \theta}{r} \; \frac{\partial S}{\partial r} \;.
\eqno(4)
$$
The $\varphi$-component of the electric current maintaining the dipolar
magnetic configuration is given by
$$
j_\varphi = - \frac{c}{4 \pi} \cdot \frac{\sin \theta}{r} \cdot \left(
\frac{\partial^{2} S}{\partial r^{2}} - \frac{2 S}{r^{2}} \right) \;.
\eqno(5)
$$
Then one has for the rate of Joule heating $\dot{q} = j^{2}/\sigma$
$$
\dot{q} = \frac{c^{2}}{16 \pi^{2} \sigma} \cdot \frac{\sin^{2} \theta}
{r^{2}} \cdot \left(\frac{\partial^{2} S}{\partial r^{2}} -  
\frac{2 S}{r^{2}} \right)^{2} \;. \eqno(6)
$$
For the sake of simplicity, we will neglect non-sphericity in cooling
calculations. Therefore, instead of equation (6), we will use the
polar-averaged expression for Joule heating,
$$
\dot{q} = \frac{c^{2}}{24 \pi^{2} r^{2} \sigma} \left(
\frac{\partial^{2} S}{\partial r^{2}} - \frac{2 S}{r^{2}} \right)^{2} \;.
\eqno(7)
$$
It is convenient to normalize the function $S(r,t)$ to its initial
value at the surface, $S(R,0)$, which in its turn can be related to the 
initial field strength at the magnetic equator, $B_{e}$, by $S(R,0) =
R^{2} B_{e}$. Finally, we obtain the expression for the rate of 
Joule heating in the form
$$
\dot{q} = \frac{c^{2} R^{4} B_{e}^{2}}{24 \pi^{2} r^{2} \sigma}
\left( \frac{\partial^{2} s}{\partial r^{2}} - \frac{2 s}{r^{2}}
\right)^{2} \;,  \eqno(8)
$$
where $s(r,t) = S(r,t)/S(R,0)$. 

The evolution of the magnetic field as well as the rate of heat
production is strongly sensitive to the conductivity. In the crust,
the conductivity is determined by electron scattering with phonons
and lattice impurities; which mechanism dominates depends on the
density $\rho$ and the temperature $T$. Electron-phonon scattering
gives the main contribution to the conductivity at a relatively
low density. For this mechanism, $\sigma \propto T^{-1}$ when $T$
is above the Debye temperature and $\sigma \propto T^{-2}$ for lower
$T$. Electron-impurity scattering becomes more important with
increasing density and decreasing $T$. When it dominates, $\sigma$ is
nearly independent of $T$, and its magnitude depends on the impurity
parameter
$$
\xi = \frac{1}{n_{i}} \sum_{n_{\alpha}} n_{\alpha}(Z-Z_{\alpha})^{2} \;,
$$
where $n_{i}$ and $Z$ are the number density and charge number of the
dominant background ion specie, respectively, and $n_{\alpha}$ is the
number density of an interloper specie of charge $Z_{\alpha}$; the
summation is over all species of impurities. In our calculations, we use 
the numerical data on the phonon conductivity obtained by Itoh et al. 
(1993) and the analytical expression for the impurity conductivity 
derived by Yakovlev \& Urpin (1980).
\begin{figure}
\centerline{\psfig{figure=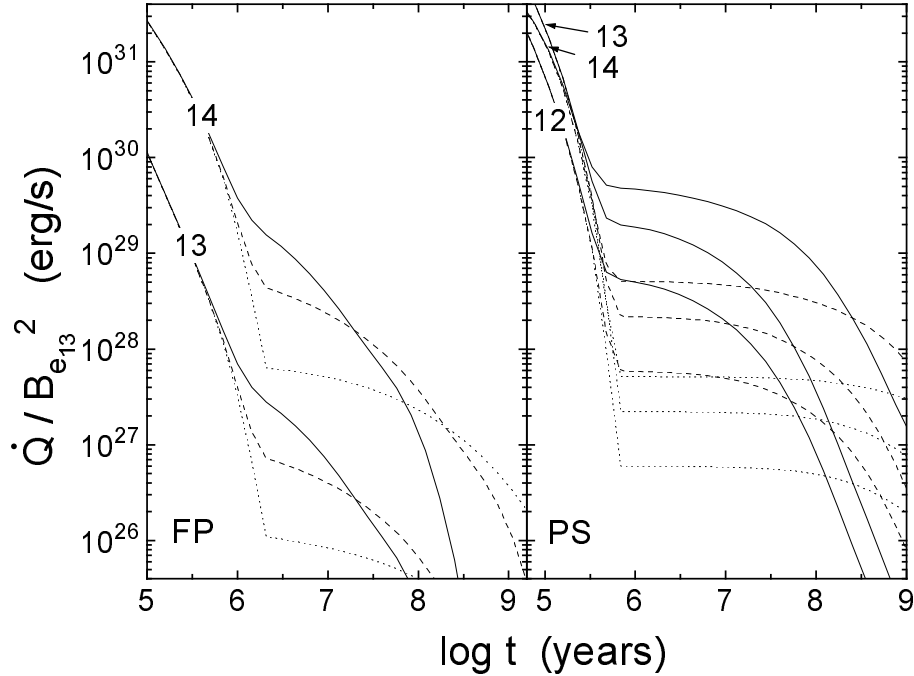,width=12cm}}
\caption{
The time dependence of the rate of Joule heating
integrated over the star volume and normalized to $B_{e13} = B_{e}/
10^{13}$ G. Numbers near the curves indicate the logarithm of $\rho_{0}$.
Different types of lines correspond to different values of $\xi$: 0.1
(solid), 0.01 (dashed), 0.001 (dotted).}
\end{figure}

Our cooling calculations taking into account the additional heating 
caused by the ohmic dissipation have been done within the {\it
isothermal approximation} that follows from assuming constant
temperature (corrected by the redshift factor) in the interior of the
star ($\rho > 10^{10}$ g/cm$^{3}$) and a temperature drop at the surface 
given by the atmosphere model that relates the interior temperature 
with the effective temperature (outer boundary condition) and allows
to calculate the photon luminosity of the star. Although the
calculations presented here correspond to the isothermal approximation 
we have checked for isothermality running some models where this
restriction is not enforced. At time when Joule heating becomes 
important, the thermal conductivity of the crust is so high that the
isothermal condition is very well satisfied and we do not observe
significant differences in the thermal evolution with respect to the 
isothermal approximation. 

We use the atmosphere models calculated by Van Riper (1988) to impose
the outer boundary condition to the cooling calculations. These 
atmosphere models are obtained by solving the hydrostatic and
radiative equilibrium equations for a pure $^{56}$Fe composition 
(see Van Riper 1988 for details).

Calculations presented here are based on the so called {\it standard 
cooling scenarios} which correspond to a star with standard neutrino
emissivities. We consider the thermal history for $1.4 M_{\odot}$
models constructed with the equations of state of Friedman 
and Pandharipande (1981; hereafter FP) and Pandharipande and Smith (1976; 
hereafter PS). The PS model is representative of stiff equations of state 
with a low central density and a massive crust; the FP model represents 
intermediate equations of state. The stiffer the equation of state, the 
larger the radius and crustal thickness for a given neutron star mass. 
For the FP and PS models, the radii are 10.61 km and 15.98 km, respectively; 
the corresponding crustal thicknesses are $\approx$980 m and 4200 m; the 
crust bottom is located at the density $2 \times 10^{14}$ g/cm$^{3}$. Our 
choice is imposed by the fact that only models with the equation of
state stiffer than that of Friedman and Pandharipande seem to be
suitable to account for the available observational data on the magnetic
evolution of pulsars (Urpin \& Konenkov 1997).

\section{Numerical results}

In our calculations, we assume the initial magnetic field to be
confined to the outer layers of the crust with density $\rho \leq
\rho_{0}$. The calculations have been performed for a wide range of
$\rho_{0}$, $10^{14} \geq \rho_{0} \geq 10^{12}$ g/cm$^{3}$. In the
present paper, we choose the initial function $s(r,0)$ in the form
$$
s(r,0) = (1-r^{2}/r_{0}^{2})/(1 - R^{2}/r_{0}^{2}) \;, \;\;\; 
r \geq r_{0}
$$
$$
s(r,0) = 0 \;, \;\;\; r < r_{0}
$$
where $r_{0}$ is the boundary radius of the region originally
occupied by the magnetic field, $\rho_{0} = \rho(r_{0})$. Note that the 
field decay and Joule heating are sensitive to the initial depth
penetrated by the field and, hence, to the value $\rho_{0}$. However,
both these quantities are much less affected by the particular form
of the original field distribution.

The impurity parameter, $\xi$, is taken within the range $0.1 \geq \xi \geq 
0.001$ and it is assumed to be constant throughout the crust. This
parameter plays a key role in any model of the magnetic evolution of
neutron stars but, unfortunately, there are no reliable theoretical
estimates of $\xi$. Calculations of the impurity charge and concentration
in the crust meet troubles because of large uncertainties in the
non-equilibrium processes during the very early evolution of a neutron
star. Flowers \& Ruderman (1977) made an attempt to estimate the final
crustal composition taking into account slow neutrino reactions at
$T < 10^{10}$K (but above the melting point). They calculated small
deviations from the equilibrium composition predicted by
energy-minimisation criteria and estimated $\xi \approx 0.004$.
However, this estimate contains the binding energies of nuclei in
the exponential factor thus the obtained value is rather uncertain.
The properties of the crust can also be influenced by accretion during
the early phase after the supernova explosion (see, e.g, Chevalier
1989). During this phase, the unbind fraction of matter falls back onto
the neutron star surface and the total amount of accreted mass may be
as large as $0.1 M_{\odot}$. Evidently, this material can substantially
change the crustal composition and increase the impurity parameter.
Besides, $\xi$ can generally be non-uniform within the crust (De Blasio
1998) because mixing between the layers with different composition may
be an important mechanism of the impurity production. Note also that
calculations of the magnetic evolution of neutron stars with the crustal
magnetic field (Urpin \& Konenkov 1997) give a better fit to
observational data if $\xi \sim 0.1-0.01$.
\begin{figure}
\centerline{\psfig{figure=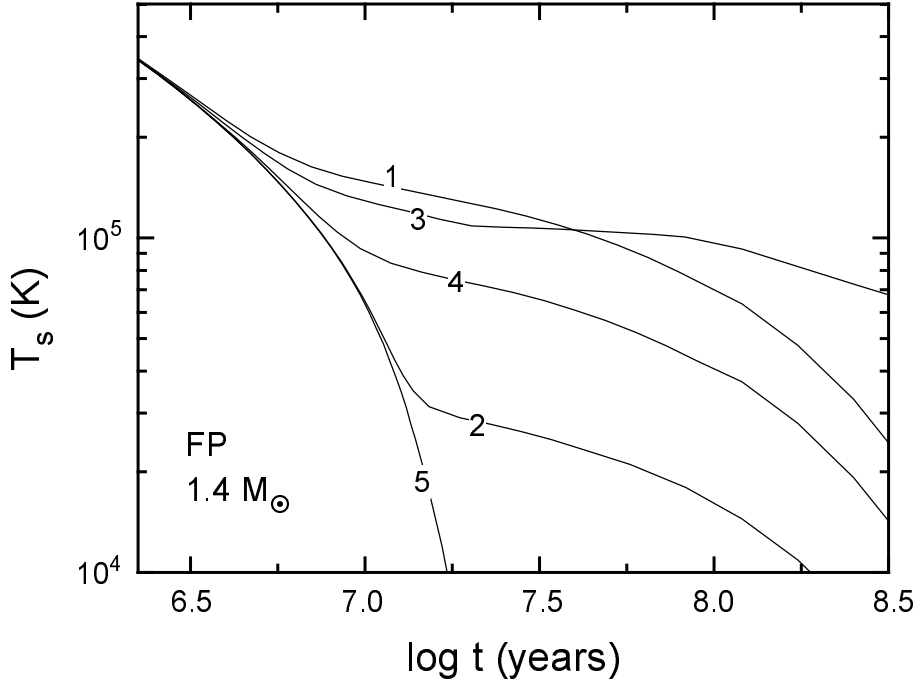,width=12.cm}}
\caption{
The thermal evolution of the FP model with and without
Joule heating; curve 1 -- $B_{e}=1.5 \times 10^{13}$ G, $\rho_{0}=
10^{14}$ g/cm$^{3}$, $\xi=0.1$; curve 2 -- $B_{e}= 1.5 \times 10^{13}$ G,
$\rho_{0}= 10^{13}$ g/cm$^{3}$, $\xi=0.1$; curve 3 -- $B_{e}=1.5 \times
10^{13}$ G, $\rho_{0}=10^{14}$ g/cm$^{3}$, $\xi= 0.01$; curve 4 --
$B_{e} = 5 \times 10^{12}$ G, $\rho_{0}=10^{14}$ g/cm$^{3}$, $\xi=0.1$;
curve 5 -- without Joule heating.}
\end{figure}

Figure 1 plots the evolution of the rate of Joule heating integrated over
the neutron star volume, 
$$
\dot{Q} = 4 \pi \int_{0}^{R} \dot{q} r^{2} dr \;,
$$
for the FP and PS models. Calculations have been performed for few values 
of $\rho_{0}$ and $\xi$. At $t> 1$ Myr, the efficiency of Joule heating 
at given $\rho_{0}$ turns out to be essentially different for the FP and 
PS models with a much lower rate of heating for the FP model. This 
difference is evident because the thickness of the crust is much smaller 
for the FP model and, hence, the field strength decreases faster. Since 
the rate of Joule heating is proportional to $j_{\varphi}^{2}$, it is much 
greater during the late evolution for the PS model which experiences a lower 
field decay. The rate of heating is very sensitive to the initial depth 
penetrated by the field. If the field is initially confined to the layers 
with density $\rho < 10^{14}$
g/cm$^{3}$ for the FP model and $\rho < 10^{12}$ g/cm$^{3}$ for the
PS model then, after $\sim 1$ Myr, the rate of heating is likely too low 
to heat the neutron star to a sufficiently high surface temperature even 
if the initial magnetic field is of the order of the maximal field 
observed in neutron stars, $\sim 4 \times 10^{13}$ G. Note that, at
$t < 1$ Myr, the effect of Joule heating on the thermal evolution is 
negligible (see below).  The rate of heating is also sensitive to the 
impurity parameter $\xi$. For a given $\rho_{0}$ and $B_{e}$, $\dot{Q}$ is 
initially higher for larger values of $\xi$ since the conductivity is lower 
for such $\xi$. However, a lower conductivity leads to a faster decrease of the
field strength and, hence, $\dot{Q}$. At some age (which generally depends 
on $\rho_{0}$ and $\xi$), the heat production becomes larger for a smaller
$\xi$. Note that the ohmic dissipation can maintain the rate of heating 
at approximately the same level during extremely long time. Thus,
for a very pure crust with $\xi=0.001$, the neutron star may have
a practically constant surface temperature during $\approx 1000$ Myr after 
the initial cooling stage ($t< 1-3$ Myr). For more polluted crust with 
$\xi=0.01$, this phase of 
almost constant surface temperature can last $\sim 100$ Myr.

Figure 2 shows the thermal evolution of the neutron star with Joule
heating and with the FP equation of state. We plot the cooling curves
only for the age $t > 2 \times 10^{6}$ yr since the earlier evolution is 
not affected practically by Joule heating. For a comparison, 
the cooling  history of a non-magnetized neutron star is also shown. All the 
cooling curves presented in this paper are obtained considering a 
non-magnetized atmosphere. This is done so for the sake of simplicity in the 
comparison of the different models. Had we used magnetic atmospheres we would 
have obtained significant differences in the evolution of the surface 
temperature with respect to the non-magnetized atmospheres only before 
the Joule heating drives the evolution but not after this time, as we 
will explain bellow.

It turns out that Joule heating may have an appreciable influence
on the thermal evolution of the FP model only if the magnetic field 
occupies initially a significant fraction of the crust volume and if 
the field is initially very strong. Therefore, calculations presented
here have been performed for $B_{e} \geq 5 \times 10^{12}$ G and $\rho_{0} 
= 10^{13}$ and $10^{14}$ g/cm$^{3}$. Note that the value $\rho_{0}=10^{14}$ 
g/cm$^{3}$ corresponds to a depth from the surface of $\approx 660$ m,  
thus, about 60\% of the crust volume has to be occupied initially by the 
field. Except a short initial phase ($\sim 3-10$ Myr) when the surface 
temperature is high, the effect of Joule heating on neutron star cooling 
turns out surprisingly simple: approximately all heat released due to 
the field dissipation has to be emitted from the surface, thus,
the surface temperature $T_{s}$ obeys with a high accuracy the equation 
$$
\dot{Q} \approx 4 \pi R^{2} \sigma_{SB} T_{s}^{4} \;,  \eqno(9)
$$
where $\sigma_{SB}$ is the Stephan-Boltzmann constant. The heat flux 
to the interior is negligible for all considered models and, therefore, 
the luminosity of old neutron stars has to be completely determined by 
the field strength and geometry and the conductive properties of the 
crust. Note that equation (9) is also valid for the PS model (see below).

If the initial field of a neutron star is of the order of the maximal
pulsar field, $B_{e} \sim (3-4) \times 10^{13}$ G, Joule heating can maintain 
a relatively high surface temperature $\sim 5 \times 10^{4} - 10^{5}$ K in
relatively old neutron stars with $t \geq 10$ Myr. Evidently, this 
temperature is much above the surface temperature predicted by the
standard cooling models without additional heating mechanisms. For the
FP model, the neutron star can stay in such a ``warm'' state rather long,
$\sim 30-60$ Myr. The efficiency of heating is very sensitive to 
$\rho_{0}$: if the field is initially anchored in the layers with 
$\rho \leq 10^{13}$ g/cm$^{3}$ the surface temperature at $t>10$ Myr
turns out to be lower than $3 \times 10^{4}$ K even for the maximal
pulsar field.
\begin{figure}
\centerline{\psfig{figure=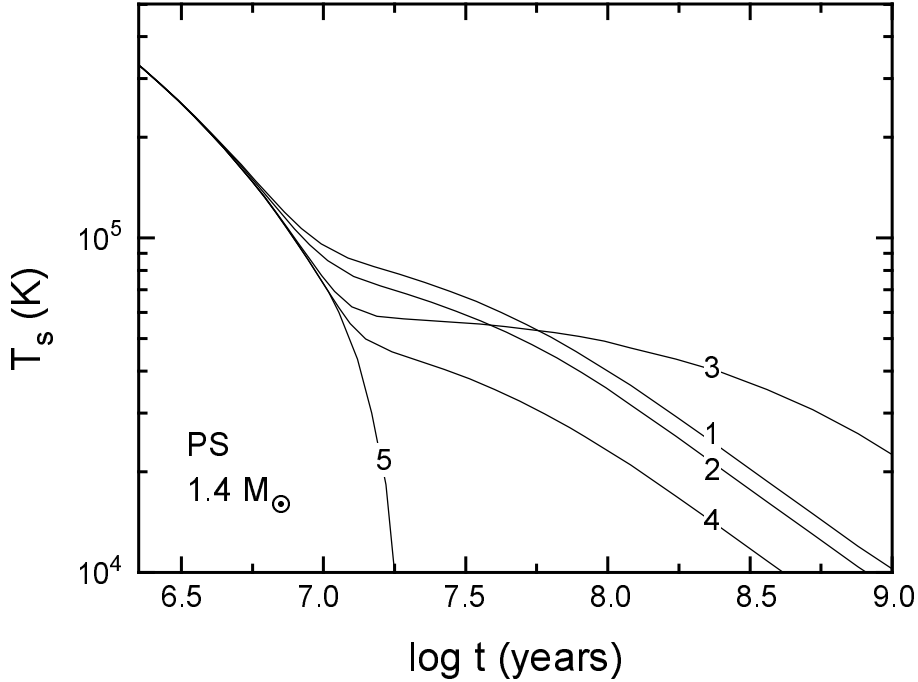,width=12.cm}}
\caption{
The thermal evolution of the PS model with and without
Joule heating; curve 1 -- $B_{e}=1.5 \times 10^{13}$ G, $\rho_{0}=
10^{13}$ g/cm$^{3}$, $\xi= 0.1$; curve 2 -- $B_{e}= 1.5 \times 10^{13}$ G,
$\rho_{0}= 5 \times 10^{12}$ g/cm$^{3}$, $\xi= 0.1$; curve 3 -- $B_{e}=
1.5 \times 10^{13}$ G, $\rho_{0}= 10^{13}$ g/cm$^{3}$, $\xi=0.01$;
curve 4 -- $B_{e}=5 \times 10^{12}$ G, $\rho_{0}=10^{13}$ g/cm$^{3}$,
$\xi=0.1$; curve 5 -- without Joule heating.}
\end{figure}

In Figure 3, we plot the thermal evolution of the PS model with and
without Joule heating. The cooling curves are shown only for $t > 
2$ Myr since for the earlier age the influence of Joule heating is
unimportant. Like in the case of the FP model, at the late evolutionary
stage ($t > 10$ Myr) the surface temperature of the PS model is 
determined by balancing the Joule heating with the photon luminosity 
(equation 9). However, for the PS model the effect of additional
heating is much more pronounced because the field decay is substantially
slower and, after $10$ Myr of evolution, the field is stronger for this
model. The surface temperature caused by Joule heating may be as high
as $3 \times 10^{4} - 10^{5}$ K even if the field occupied initially
a relatively small fraction of the crust volume. It turns out that 
$T_{s}$ is strongly sensitive to all parameters determining the magnetic 
evolution ($\xi$, $\rho_{0}$, $B_{e}$). At $t>10$ Myr, a relatively high 
temperature ($T_{s} \geq 5 \times 10^{4}$ K) can be reached only for strongly 
magnetized neutron stars with the initial field $B_{e} \geq 5 \times 
10^{12}$ G. The surface temperature is rather low if the magnetic field is 
initially confined to the layers with a small density, $\rho \leq 
10^{12}$ g/cm$^{3}$. This is due to the fact that the crustal field anchored 
initially in not very deep layers experiences a fast decay during the 
very early evolutionary stage ($t < 10^{5}$ yr) when the neutron star 
is hot and the crustal conductivity is low. For such initial magnetic 
configurations, the field strength at $t > 10$ Myr is too weak to
produce an appreciable Joule heating. The most remarkable point is that
all considered models with Joule heating can
maintain a sufficiently high temperature during extremely long
time. Thus, our calculations show that for the ``polluted'' crust
($\xi=0.1$) $T_{s}$ decreases only by a factor $\sim 3$ when $t$ runs from
10 to 100 Myr. For $\xi=0.01$, the temperature is practically
unchanged during the same period. Evidently (see equation 9), the 
characteristic cooling time in our model is determined by the ohmic
decay time of the magnetic field and, therefore, should be very long.
\begin{figure}
\centerline{\psfig{figure=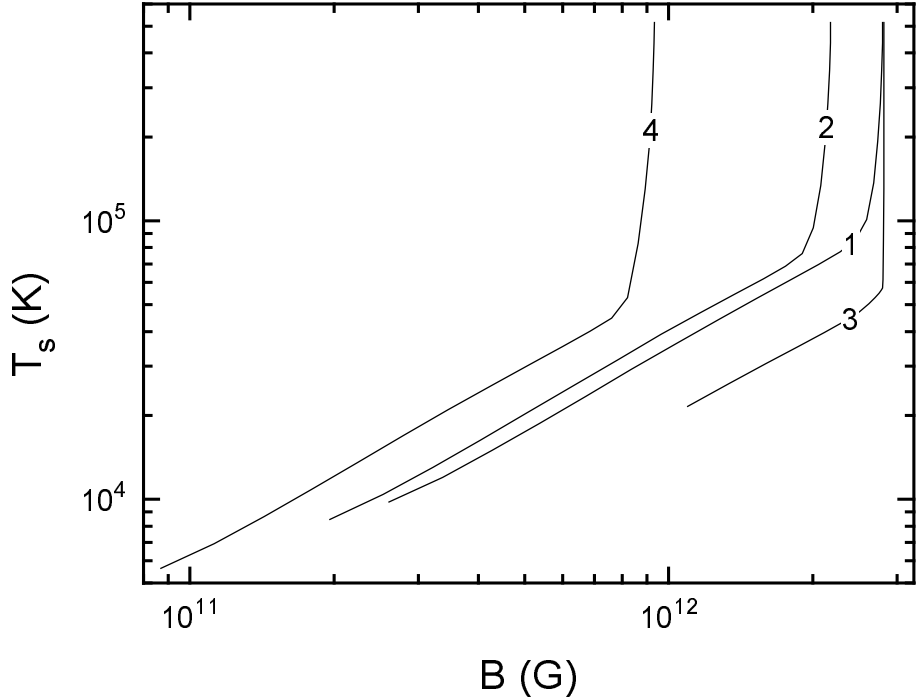,width=12.cm}}
\caption{
The dependence of the surface temperature, $T_{s}$,
on the current equatorial magnetic field, $B$, for the PS model. Numbers
near the curves correspond the same values of parameters as in Fig.3.}
\end{figure}

Note that a simple estimate of the surface temperature caused by Joule
heating can be obtained directly from equation (9). The rate of heat
production is equal to the rate of decrease of the magnetic energy. In the
main fraction of the crust volume (except surface layers and a region
near the magnetic pole), $\theta$-component of the crustal field is
stronger than the radial one, which in its turn is of the order of the
current surface field, $B(t)$. We have $B_{\theta} \sim B_{r} (R/\ell)
\sim B(t) (R/\ell)$ (see equation (4)) where $\ell$ is the radial
lengthscale of the magnetic field; this lengthscale depends on time
since the field diffuses inwards. Therefore, the energy of the crustal
field can be estimated as $E_{m} \sim 4 \pi R^{2} \ell (B_{\theta}^{2}/8
\pi)$ and, correspondingly, $\dot{Q} \sim E_{m}/t$. Substituting this
expression into equation (9), we obtain the estimate of the surface
temperature,
$$
T_{s} \sim \left( \frac{R^{2} B^{2}(t)}{8 \pi \sigma_{SB} \ell t}
\right)^{1/4},
$$
or
$$
T_{s} \sim 4 \times 10^{4} B_{12}^{1/2}(t) R_{6}^{1/2} \ell_{5}^{-1/4}
t_{8}^{-1/4} {\mathrm K}, \eqno(10)
$$
where $B_{12}(t) = B(t)/10^{12}$G, $R_{6}=R/10^{6}$cm, $\ell_{5}=
\ell/10^{5}$cm, and $t_{8}=t/10^{8}$yr. In this equation, both the
current field strength, $B(t)$, and the depth penetrated by the
field, $\ell$, depend generally on the impurity parameter, $\xi$, and
these dependences are rather complex because of a non-uniform
chemical composition of the crust. Nevertheless, sometimes, the
estimate (10) may be useful because the dependence of $T_{s}$ on
$\ell$ is weak, and $\ell$ varies within a relatively narrow range. For
example, for the models presented here, $\ell_5$ is ranged from 0.5 to
1 for the FP model and from 1 to 4 for the PS model.
Of course, this simplified estimation is only valid soon after the Joule
heating becomes important $t \sim$ 10--30 Myr  till the magnetic field
reaches the crust--core boundary ($t \sim$ 1000 Myr for the PS model).

Obviously, in the suggested mechanism, the rate of Joule heating and 
the surface temperature depend strongly on the current magnetic field 
strength at the equator, $B$. In Figure 4, we plot the dependence of the 
surface temperature on $B$ for the PS neutron star model. The chosen range 
of the field strength corresponds to the age from $\sim 3$ Myr to $\sim 
1000$ Myr (see Fig.3). Note that the  magnetic field strength, plotted 
in Fig.4, is different from what is usually calculated using observational
data on the pulsar period, $P$, and its derivative, $\dot{P}$, and assuming 
magnetodipole braking. The standard estimate (Ostriker \& Gunn 1969)
gives the field strength at the magnetic equator, 
$$
B_{obs}= (3 I c^{3} P \dot{P}/ 8 \pi^{2} R^{6})^{1/2} \;, \eqno(11)
$$ 
where $I$ is the moment of inertia. However, this equation gives an 
estimate of the equatorial field produced by the component of the
magnetic dipole perpendicular to the spin axis because the parallel
component does not contribute to braking. The rate of Joule
heating is determined by the true magnetic field produced by the both 
components. Therefore, the equatorial field strength entering our
calculations is by a factor $1/\sin \alpha$ larger than the observable
pulsar magnetic field, $B_{obs}$, where $\alpha$ is the angle between 
magnetic and spin axes. 

To plot $T_{s}(B)$, we simply eliminate the $t$-dependence 
from the cooling curves, $T_{s}(t)$, and the magnetic decay curves, $B(t)$, 
calculated for the same values of initial parameters (for more details 
concerning magnetic decay curves see Urpin \& Konenkov 1997). Obviously, 
$T_{s}$ decreases with the magnetic field decay. During the first
several Myr while Joule heating is negligible, the magnetic evolution is 
going much slower than the thermal one because neutrino emissivity and 
photon luminosity make the characteristic cooling time much shorter than 
the characteristic time for ohmic decay. This phase of evolution 
(represented by nearly vertical pieces of curves) lasts while the total 
rate of Joule heating, $\dot{Q}$, is smaller than the luminosity. At $B \leq 
(1-3) \times 10^{12}$ G (depending on the parameters) when Joule heating 
plays a dominating role in the thermal evolution, the characteristic 
cooling time becomes comparable with the decay time of the magnetic field 
as it follows from equation (9).

Figure 5 shows the dependence of the surface temperature on the spin-down 
age of neutron stars with the PS equation of state. Our knowledge of 
the field strength and its behaviour with time comes mainly from radio 
pulsars with measured spin-down rates. For the most of pulsars, the true 
age is unknown and observations provide information only on the so 
called spin-down age, $\tau = P/2 \dot{P}$. Therefore, for a comparison 
with observational data, it is convenient to analyse the 
dependence of $T_{s}$ on $\tau$ rather than on $t$. Assuming magnetodipole
braking and integrating equation (11), we can calculate the spin-down
age as a function of time, $\tau = \tau(t)$. Eliminating $t$ from the couple
of functions $T_{s}(t)$ and $\tau(t)$, we obtain the dependences 
$T_{s}(\tau)$ shown in Figure 5. 
\begin{figure}
\centerline{\psfig{figure=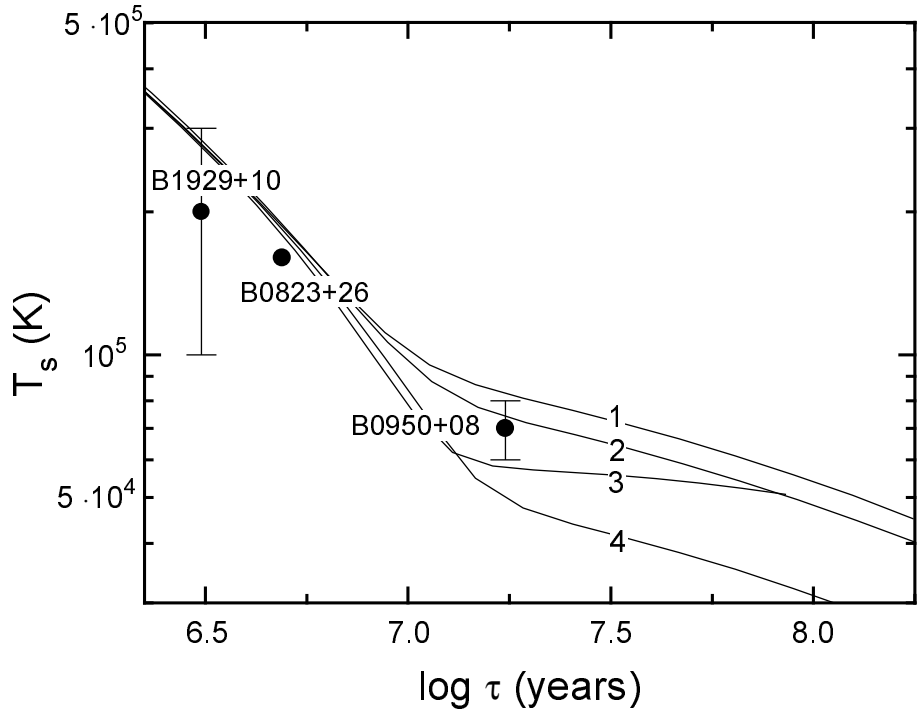,width=12.cm}}
\caption{
The dependence of the surface temperature, $T_{s}$,
on the spin-down age, $\tau$, for the PS model. Numbers near the curves
correspond the same values of parameters as in Fig.3.}
\end{figure}

The $\tau$-dependence of cooling curves is qualitatively similar to 
their $t$-dependence. The only difference is a bit slower decrease of
$T_{s}$ in terms of the spin-down age. This difference is clear because
$\tau(t) > t$ for a decaying magnetic field. In Figure 5, we also plot
the data for three middle age and old pulsars: B0823+26 
($\tau = 4.9$ Myr), B1929+10 ($\tau=3$ Myr) and B0950+08 ($\tau =
17.4$ Myr). The surface temperatures of these pulsars are $\sim 1.6 
\times 10^{5}$ (this estimate has been obtained from the luminosity
given by Becker \& Tr\"umper, 1997, assuming the black-body spectrum), 
$(1-3) \times 10^{5}$ and $(7 \pm 1) \times 10^{4}$K (Pavlov, 
Stringfellow \& Cordova 1996), respectively. Of course, these 
observational data are too poor to infer somewhat categorical but, 
nevertheless, it seems that the ohmic dissipation can produce enough
heat to maintain the observed surface temperatures of middle age and
old pulsars.

\section{Summary}

We considered Joule heating caused by the decay of the crustal
magnetic field in neutron stars. Calculations of the thermal evolution
of neutron stars show that the heat released in the crust due to
the field decay diffuses mainly outward thus practically all the Joule 
heat has to be radiated from the surface. Due to this, the surface
temperature at the late evolutionary stage ($t> 10$ Myr) turns out to
be independent of the atmosphere models and is determined by balancing
between the rate of Joule heating integrated over the neutron star 
volume and the luminosity (see equation (9)). Being independent of
the atmosphere models, $T_{s}$ is however strongly dependent on 
parameters of the magnetic configuration and the conductive properties 
of the crust. Therefore, the observational study of the late thermal
history of neutron stars could be a useful diagnostic of their internal 
magnetic fields and properties of the crust. 

The decay of the crustal magnetic field can produce enough heat to
maintain a sufficiently high surface temperature $\sim 3 \times 10^{4} -
10^{5}$K. Our calculations predict that Joule heating becomes important
after a relatively short ($\sim 3-10$ Myr depending on the model) initial 
phase when the neutron star cools down to $T_{s} \sim 3 \times 10^{4} -
10^{5}$K. The further thermal evolution slows down substantially:
a characteristic cooling time becomes comparable with the decay time of the 
magnetic field. Since the field decay in the crust is very slow, the
neutron star can maintain a surface temperature practically unchanged
during extremely long time, $t \geq 100$ Myr.

\section*{Acknowledgement}

This work has been supported in part by the Spanish DGICYT (grant PB94-0973) 
and by the Russian Foundation of Basic Research (grant 97-02-18086).
It is a pleasure to thank Ken Van Riper for useful discussions.

\newpage

\section*{References}

Becker, W., \& Tr\"umper, J. 1998, A\&A (in press) \\
Bhattacharya, D., Wijers, R., Hartman, J., \& Verbunt, F. 1992, A\&A,
254, 198 \\
Chevalier, R. 1989, ApJ, 346, 847 \\
De Blasio, F. 1998, MNRAS (in press) \\
Flowers, E., Ruderman, M. 1977, ApJ, 215, 302 \\
Friedman, B., Pandharipande, V.P. 1981, Nucl.Phys. A, 361, 502 \\
Itoh, N., Hayashi, H., Kohyama, Y. 1993, ApJ, 418, 405 \\
Mignani, R., Caraveo, P.A., \& Bignami, G. 1997, ApJ, 474, L51 \\
Narayan, R., \& Ostriker, J.P. 1990, ApJ, 352, 222 \\
Nomoto, K., \& Tsuruta, S. 1987, ApJ, 312, 711 \\
Ostriker, J.P., \& Gunn, J.E. 1969, ApJ, 157, 1395 \\
Pandharipande, V.R., Pines, D., Smith, R.A. 1976, ApJ, 208, 550 \\
Pavlov, G., Stringfellow, G.S., \& Cordova, F.A. 1996, ApJ, 467, 370 \\
Sang, Y., \& Chanmugam, G. 1987, ApJ, 323, L61 \\
Schaaf, M.E. 1990, A\&A, 227, 61 \\
Shibazaki, N., \& Lamb, D. 1989, ApJ, 346, 808 \\
Umeda, H., Shibazaki, N., Nomoto, K., \& Tsuruta, S. 1993, ApJ, 408, 186 \\
Urpin, V., \& Konenkov, D. 1997, MNRAS, 292, 167 \\
Urpin, V., Geppert, U., \& Konenkov, D. 1998, MNRAS (in press) \\
Van Riper, K. 1988, ApJ, 329, 339 \\
Van Riper, K. 1991, ApJS, 75, 449 \\
Vidaurre, A., P\'erez, A., Sivak, H., Bernab\'eu, J., \& Ib\'a\~nez, J.M.
1995, ApJ, 448, 264 \\
Yakovlev, D., \& Urpin, V. 1980, SvA, 24, 303 \\

\end{document}